\documentclass[
    10pt,
    twocolumn,
    final,
    superscriptaddress,
    prl,
    showpacs
]{revtex4}

\usepackage{graphicx}
\usepackage{amssymb}
\usepackage{amsmath}

\newcommand{\intd}{d}
\newcommand{\intR}{\int_{-\infty}^\infty}

\newcommand{\etal}{\mbox{\textit{et al.}}}

\newcommand{\uev}{$\mu$\textrm{eV}}
\newcommand{\mev}{\textrm{meV}}

\newcommand{\Eqref}[1]{Eq.~(\ref{#1})}
\newcommand{\Figref}[1]{Fig.~\ref{#1}}

\begin{document}
\title{Inelastic Scattering and Local Heating in Atomic Gold
Wires}
\date{\today}
\eprint{\mbox{This work is published in Phys.~Rev.~Lett. {\bf 93}, 256601 (2004)}. }
\addtolength{\textheight}{5mm}


\author{Thomas \surname{Frederiksen}}
\email{thf@mic.dtu.dk}
\author{Mads \surname{Brandbyge}}
\affiliation{MIC -- Department of Micro and Nanotechnology,
Technical University of Denmark, {\O}rsteds Plads, Bldg.~345E,
DK-2800 Lyngby, Denmark}
\author{Nicol\'{a}s \surname{Lorente}}
\affiliation{Laboratorie Collisions, Agr\'{e}gats,
    R\'{e}activit\'{e}, IRSAMC, Universit\'{e} Paul Sabatier,
    118 Route de Narbonne, F-31062 Toulouse, France}
\author{Antti--Pekka \surname{Jauho}}
\affiliation{MIC -- Department of Micro and Nanotechnology,
Technical University of Denmark, {\O}rsteds Plads, Bldg.~345E,
DK-2800 Lyngby, Denmark}


\begin{abstract}
  We present a method for including inelastic scattering
  in a first-principles density-functional computational scheme for molecular
  electronics.  As an application, we study two geometries of
  four-atom gold wires corresponding to two different values of strain,
  and present results for nonlinear differential conductance vs device bias. Our theory is in
  quantitative agreement with experimental results, and explains
  the experimentally observed mode selectivity. We also identify
  the signatures of phonon heating.
\end{abstract}

\maketitle


Atomic-size conductors are the components of the emerging
molecular electronics~\cite{Agrait&Yeyati&Ruitenbeek.2003.QPASC}.
The corresponding molecular devices have new functionalities that
exploit quantum phenomena, such as phase coherence and resonances.
A substantial effort has been devoted to molecular electronics,
producing a wealth of experimental data on electronic transport at
the molecular level,
e.g.,\cite{ReedEtAl.1997.CMJ,{SmitEtAl.2002.MCHM},{KubatkinEtAl.2003.SEToaSOMwatSRS}}.
Most recently the issue of vibrational effects has drawn much
attention since inelastic scattering and energy dissipation inside
atomic-scale conductors are of paramount importance in device
characteristics, working conditions, and -- especially --
stability~\cite{KuLaPa04,WaLeKr04,Smit&Untiedt&Ruitenbeek.2004.HSMC}.

Inelastic effects are interesting, not only because of their
potentially detrimental influence on device functioning, but also
because they can open up new possibilities and operating modes.
Indeed, these effects have been used to identify the vibrational
spectra of objects in tunneling junctions. This is the case of the
inelastic electron tunneling spectroscopy (IETS) both in
metal-insulator-metal junctions~\cite{Hansma} and on surfaces with
the scanning tunneling microscope
(STM)~\cite{Stripe&Rezaei&Ho.1998.SMVSM}. Recently, similar
vibrational signatures in the high-conductance regime have been
revealed~\cite{SmitEtAl.2002.MCHM,
AgraitEtAl.2002.OfEDiBAW,Agrait&UntiedtEtal.2002.ETPAW}. In one of
these studies Agra\"{\i}t and co-workers used a cryogenic STM to
create a free-standing atomic gold wire between the tip and the
surface of the substrate. The STM was then used to measure the
conductance against the displacement of the tip, making it
possible to determine the approximate size as well as the level of
strain of the wire. The data show distinct drops of conductance at
particular tip-substrate voltages (symmetric around zero bias),
consistent with the interpretation that the conducting electrons
were backscattered from vibrations. It was assumed that that the
onset of the drops coincided with a natural frequency of the wire
at certain sizes and strains.

Several different theories have been put forward to address the
effects of vibrations on electrical conductance. In the tunneling
regime a substantial theoretical effort was undertaken right after
the first experimental \mbox{evidence~\cite{Lambe&Jaklevic.1968}}
of vibrational signals in the tunneling
conductance~\cite{Appelbaum,Caroli}. Later, general tight-binding
methods including inelastic effects were
developed~\cite{Bonca,Emberly&Kirczenow.2000.LTISaETiMW}. More
recently, the combination of {\it ab initio} techniques (such as the
density-functional theory, DFT), and nonequilibrium Green's
function (NEGF) techniques led to a microscopic understanding of
conduction processes in the {\it elastic} regime, e.g.,
\cite{BrandbygeEtAl.2002.DFMfNEET}. Detailed {\it ab initio} studies of
IETS with STM have also appeared~\cite{Mingo,Lorentea}. To the
best of our knowledge, only few realistic calculations have
addressed inelastic effects in the high-conductance regime.
Montgomery and
co-workers~\cite{MontgomeryEtAl.2003.ICVSoAW,Montgomery&Todorov.2003.EPIAS}
used a lowest order perturbation theory (LOPT) approach for the
electron-phonon (\mbox{e-ph}) interaction to estimate the
inelastic contribution to the current through atomic gold wires
within a tight-binding description. LOPT have also been combined
with {\it ab initio} methods to study vibrational effects in point
contacts and molecular
junctions~\cite{Alavi&Taylor&GuoEtAl.2001.CTVE,ChenZwolakDiVentra.2004.ICVAMJ}.
LOPT cannot be applied in all circumstances; a point in case is
polaronic effects which have been shown to be essential for the
correct description of transport in long
chains~\cite{Ness&Shevlin&Fisher.2001.CEPH}. Unfortunately, going
beyond LOPT is a highly nontrivial task; see, e.g.,
\cite{Flensberg.2003.TBVSMT,THF_thesis,Galperin}.


In this Letter we formulate a first-principles theory of electron
transport including inelastic scattering due to phonons. We apply
it to atomic gold wires, for which high quality experimental data
are available, thus allow-ing a stringent test of the predictive
power of our scheme. We employ DFT
\cite{Perdew&Burke&Ernzerhof.1996.GGA} for the electronic
structure combined with an NEGF calculation of the steady current
and power flow. We go beyond LOPT using the self-consistent Born
approximation (SCBA) for the \mbox{e-ph} interaction. For gold
wires we find that the only significant inelastic scattering
mechanism is due to longitudinal modes with ``alternating bond
length'' (ABL) character, and show how ``heating'' of these active
modes can be identified in a transport measurement.  The
theoretically computed values for conductance changes, frequency
shift with elongation, and slope in conductance with voltage are
in excellent agreement with experiments. The theory further shows
that as the wire is stretched new vibrational modes become
effective.


Our method consists of essentially three consecutive steps
comprising the calculation of (i) mechanical normal modes and
frequencies, (ii) electronic structure and \mbox{e-ph} couplings
in a localized atomic-orbital (AO) basis set, and (iii) inelastic
transport with NEGF. We partition the system into left ($L$) and
right ($R$) electrode, and central device region ($C$), in such a
way that the direct coupling between the electrodes is negligible.
Hence we may write the electronic Hamiltonian as
\begin{equation}
{\cal H} = {\cal H}_L + {\cal V}_{LC} + {\cal H}_C(Q) + {\cal
V}_{RC} + {\cal H}_R \,, \label{eq:hamLRC}
\end{equation}
where ${\cal H}_\alpha$ is a one-electron description of electrode
$\alpha=L/R$ and ${\cal V}_{\alpha C}$ the coupling between
$\alpha$ and $C$. The central part ${\cal H}_C(Q)$ depends
explicitly on a $3N$-dimensional displacement variable $Q$ which
corresponds to mechanical degrees of freedom of $N$ atoms in
region $C$.

To obtain the most accurate normal modes $Q_\lambda$ and
frequencies $\Omega_\lambda$ within DFT of a given structure we
employ a plane-wave (PW) basis~\cite{MethodDetailsPW}. Except for
this purpose we use DFT with an nonorthogonal basis set of
numerical AOs with finite
range~\cite{SolerEtAl.2002.SIESTA,BrandbygeEtAl.2002.DFMfNEET,MethodDetailsAO},
which unambiguously allow us to partition the system as mentioned
above. In this basis we expand the $Q$-dependence of the central
part Hamiltonian to first order in $Q_\lambda$ (since the
vibrational amplitudes are small compared with the bond lengths),
and write
\begin{equation}
\mathbf{H}_C({Q}) \approx \mathbf{H}_C(0) + \sum_{\lambda=1}^{3N}
\mathbf{M}^\lambda (b_\lambda^\dagger + b_\lambda^{\mbox{}}),
\end{equation}
where $b_\lambda^\dagger$ $(b_\lambda^{\mbox{}})$ is the creation
(annihilation) operator of oscillator mode $\lambda$, and the
coupling matrices $\mathbf{M}^\lambda$ are calculated using finite
differences~\cite{HeadGordon&Tully.1992.VRMS}. If the central
region $C$ is sufficiently large the coupling elements are
localizable within its subset of the AO basis.

The transport calculation is based on  NEGF techniques and the
\mbox{e-ph} interaction treated within
SCBA~\cite{Haug&Jauho.1996.QKIOS,THF_thesis,Galperin}. The
electrical current $I_\alpha$ and the power transfer $P_\alpha$
\emph{to} the device (per spin) from lead $\alpha$ are
~\cite{Meir&Wingreen.1992.LFftCtaIEC,THF_thesis}
\begin{eqnarray}
\label{eq:currentExpr} I_\alpha &=& e\langle \dot {\mathcal
N}_\alpha\rangle =
\frac {-e}\hbar \intR \frac{\intd \omega}{2\pi} t_\alpha(\omega),\\
\label{eq:powerExpr} P_\alpha &=& -\langle \dot {\mathcal
H}_\alpha\rangle = \frac 1\hbar \intR
\frac{\intd \omega}{2\pi}\omega t_\alpha(\omega), \\
t_\alpha(\omega)&=&\textrm{Tr}[\mathbf
\Sigma^{<}_\alpha(\omega)\mathbf G^{>}(\omega) - \mathbf
\Sigma^{>}_\alpha(\omega) \mathbf G^{<}(\omega)],
\end{eqnarray}
where $\cal N_\alpha$ is the electronic number operator of lead
$\alpha$, $\mathbf G^\lessgtr$ the electronic lesser/greater
Green's function in the device region $C$, and $\mathbf
\Sigma_\alpha^\lessgtr$ the lesser/greater self-energy due to
coupling of $C$ to $\alpha$. We evaluate the SCBA e-ph self-energy
$\mathbf\Sigma_\textrm{ph}$ using free phonon Green's functions,
which involve average mode occupations $N_\lambda$ (also in
nonequilibrium). The coupled equations for $\mathbf G$ and
$\mathbf\Sigma_\textrm{ph}$ are iterated until self-consistency is
achieved. This approximation is reasonable for a weakly
interacting system as long as the mode damping rates are orders of
magnitude smaller than the oscillator frequencies. The SCBA scheme
guarantees current conservation,
i.e.~$I_L=-I_R$~\cite{THF_thesis}.

\begin{figure}[b!]
  \centering
  \includegraphics[width=.4\textwidth]{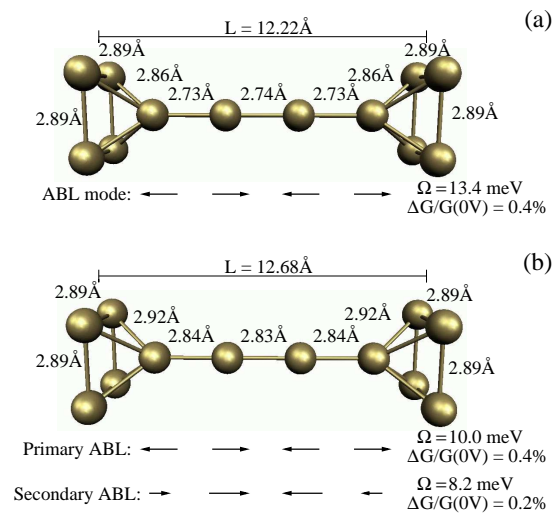}
  \caption{Geometry of a four-atom gold wire under two different states of stress
  corresponding to an electrode separation of \textsf{(a)} \mbox{$L=12.22$ \AA}
  and \textsf{(b)} \mbox{$L=12.68$ \AA}. The electrodes are
  modelled as perfect (100) surfaces, from which only the atoms closest to the wire are shown.
  The ABL modes, which cause the inelastic scattering, are shown schematically with
  arrows below each structure, together with mode energy $\Omega_\lambda$ and extracted conductance
  drop $\Delta G/G(0\textrm{V})$.}
  \label{fig:geometry}
\end{figure}

We study a linear four-atom gold wire under two different states
of strain, as shown in \Figref{fig:geometry}, corresponding to
electrode separations of \mbox{$L=12.22$ \AA} and \mbox{$L=12.68$
\AA}. The semi-infinite gold electrodes are modelled as perfect
(100) surfaces in a $3\times 3$ unit cell. We take the electrode
temperature to be $T=4.2$ K as in the experiments. Allowing the
wire atoms to move we calculate the phonon modes and energies for
each of the two structures. In the AO basis we determine the
static Hamiltonian of the whole system as well as the \mbox{e-ph}
couplings. These are then downfolded on the basis of the four wire
atoms (which constitutes region $C$) with self-energies
$\mathbf\Sigma_\alpha$ to represent the electrodes. We calculate
the phonon signal in the non-linear differential conductance vs
bias voltage ($G-V$) with \Eqref{eq:currentExpr} for two extremal
cases: the energy transferred from the electrons to the vibrations
is either (i) instantaneously absorbed into an external heat bath,
or (ii) accumulated and only allowed to leak via electron-hole
(e-h) pair excitations. We will refer to these limits as the
externally damped and externally undamped cases, respectively.

\begin{figure}[b!]
  \centering
  \includegraphics[width=.32\textwidth]{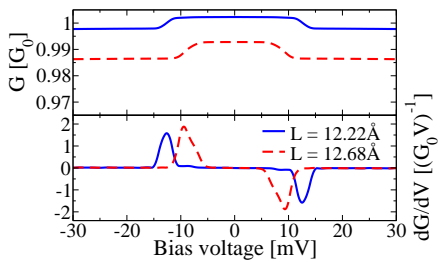}
  \caption{Differential conductance and its derivative for the four-atom gold wire at two different
  tensions in the case where the oscillators are externally damped ($N_\lambda\approx 0$).
  All modes are included in this calculation.}
  \label{fig:AllModesAndDropVsIndex}
\end{figure}

The externally damped limit corresponds to each mode having a
fixed occupation $N_\lambda\approx 0$ as set by a Bose-Einstein
distribution with a temperature $T=4.2$ K. This leads to the
results shown in \Figref{fig:AllModesAndDropVsIndex}. The
conductance is close to the quantum unit $G_0=2e^2/h$ for zero
bias and displays symmetric drops for finite bias. A comparison of
the two structures indicates that \mbox{straining} the wire
results in lower zero-bias conductance (related to weakened
couplings to the electrodes) as well as mode softening and
enhanced phonon signal. These three effects were also observed
experimentally (the shift in zero-bias conductance being most
dramatic close to rupture). The total conductance drops $\Delta
G/G(V=0)$ are found to be $0.5{\%}$ for the wire \mbox{$L=12.22$
\AA} and $0.7\%$ for \mbox{$L=12.68$ \AA}. These drops occur at
threshold voltages corresponding to the ABL mode energies. By
including one mode at a time, we can investigate the contribution
from each mode separately. This reveals that the inelastic
scattering, for both geometries, originates only from longitudinal
modes with ABL character. For the linear gold wire the conduction
channels are rotationally invariant, hence they cannot couple to
transverse modes. On the other hand for a zigzag conformation,
which under certain strains is favorable
\cite{SanchezPortalEtAl.1999.SMGW}, also transverse modes could
possibly contribute. Indistinctness of such signals are thus fully
compatible with a linear geometry. The importance of ABL character
can be understood as a reminiscence of the momentum conservation
in infinite one-dimensional wires, where the only allowed
inelastic (intraband) transitions correspond to electrons
interacting with phonons with a wavenumber of approximately twice
the Fermi wavevector
(backscattering)~\cite{Agrait&UntiedtEtal.2002.ETPAW}. For
\mbox{$L=12.22$ \AA} we find a conductance drop $\Delta G/G(V=0)$
from the ABL mode of $0.4\%$, and for \mbox{$L=12.68$ \AA} drops
of $0.4\%$ and $0.2\%$ from the primary and secondary ABL mode,
respectively. These modes and their contributions to the
conductance are also shown in \Figref{fig:geometry}. The
contribution from any other mode is found to be less than
$0.06\%$.

The salient features of the
experiments~\cite{AgraitEtAl.2002.OfEDiBAW,Agrait&UntiedtEtal.2002.ETPAW},
{\it viz.} (i) the order of magnitude of the conductance drop,
(ii) the mode softening, and (iii) the increased phonon signal
with strain, are all properly reproduced by our calculations. In
particular, we find the same frequency shift with elongation
\mbox{($\Delta \Omega/\Delta L = -7~\mev/${\AA})} as observed
experimentally. From our analysis we conclude that the enhanced
signal with strain is not due to increased \mbox{e-ph} couplings,
but rather due to the fact that the electronic structure changes.
This change affects the bond strengths, and hence the normal modes
of the structure, such that a second mode acquires ABL character.
This is contrary to considerations based on an infinite
one-dimensional wire model~\cite{Agrait&UntiedtEtal.2002.ETPAW}.

\begin{figure}[b!]
  \centering
  \includegraphics[width=.35\textwidth]{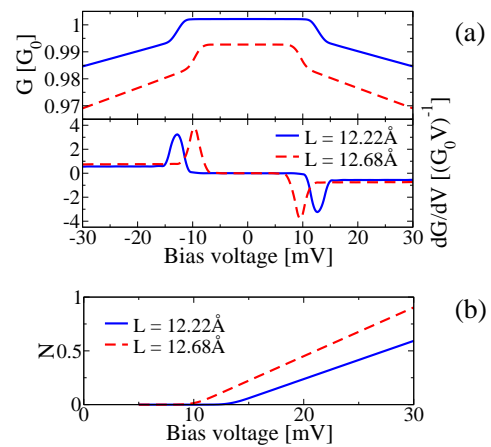}
  \caption{\textsf{(a)} Differential conductance and its derivative for the four-atom gold wire at two different
  tensions in the externally undamped limit. Only the most important mode is included in
  this calculation.
  \textsf{(b)} Mode occupation $N$ vs bias voltage.}
  \label{fig:OneModeAndHeating}
\end{figure}

In the externally undamped limit we determine the mode occupations
for a given bias voltage using the fact that the system is in a
steady state. With \Eqref{eq:powerExpr} we require that the net
power into the device \mbox{$P_L+P_R$}, which equates the net
power transferred from the electrons to the phonons, must be zero.
This in turn puts a restriction on $N_\lambda$. For simplicity we
include only the most important mode. The conductance calculation
is shown in \Figref{fig:OneModeAndHeating}a. Compared with the
externally damped results \Figref{fig:AllModesAndDropVsIndex}, the
notable differences are a slightly larger drop as well as a finite
slope in the conductance beyond the onset of inelastic scattering.
\Figref{fig:OneModeAndHeating}b shows where the vibrational
excitation sets in and starts to increase linearly with bias. At a
voltage \mbox{$V=55$ mV} the occupation is found to be the same as
\emph{if} the mode was occupied according to a Bose-Einstein
distribution with \mbox{temperature $T=300$ K.}

A finite slope was also observed in the experiments, and
speculated to be directly related to nonequilibrium phonon
populations~\cite{Agrait&UntiedtEtal.2002.ETPAW}. This is
confirmed by our calculations. Quantitatively we find $\intd
G/\intd V (20\textrm{mV}) \approx -0.6\,(G_0\textrm{V})^{-1}$ and
$\intd G/\intd V(20\textrm{mV}) \approx -0.7\,
(G_0\textrm{V})^{-1}$ for {$L=12.22$ \AA} and {$L=12.68$ \AA},
respectively, which is only slightly larger than detected for
relatively long gold wires. In reality the phonon modes are damped
also by mechanical coupling to bulk phonons in the electrodes.
This coupling depends strongly on the nature of the
chain-electrode contact and hence, understood poorly. We expect
that the typical damping conditions lead to $G-V$ curves in
between \Figref{fig:AllModesAndDropVsIndex} and
\Figref{fig:OneModeAndHeating}a.

The observed linewidth of the phonon signal is set by a
combination of both electronic temperature and mode
broadening~\cite{Hansma}. The temperature broadening alone is of
the order $5k_BT\approx 2$ {\mev} (FWHM). As the atomic wire is
elongated, new modes contribute to the drop. Hence, our
calculations show that the corresponding linewidth will increase
from 2 to 4 {\mev} due to the appearance of a second mode,
cf.~\Figref{fig:AllModesAndDropVsIndex}. In addition to this, mode
broadening due to coupling to the electrons and to vibrations in
the bulk also contribute. We estimate the damping of the modes
from e-h pair generation to be no more than
\mbox{$\gamma_\textrm{e-h}=30$-35
\uev}~\cite{Hellsing&Persson.1984}, which is thus negligible here.
In the experiment the linewidth is typically around 5 \mev, and
hence it is either a result of the overlap of several vibrational
modes or due to significant coupling to bulk modes. This could be
clarified with measurements at even lower temperatures, where it
might be possible to resolve several modes as a function of the
wire strain.

As we show elsewhere~\cite{THF_thesis,Frederiksen.2004.IWCE}, it
is possible to describe the system qualitatively with a
single-orbital tight-binding model. Using this simplified approach
longer chains can be examined, for which first-principles
calculations are not feasible at the present stage.  The simple
model predicts that the conductance drop $\Delta G/G(V=0)$ and
slope $\intd G/\intd V$ beyond the threshold scale linearly with
the number of atoms in the wire (we considered up to 40 atoms).
This supports the notion that the inelastic scattering occurs
inside the wire itself.


In conclusion, we investigated inelastic effects in atomic gold
wires using a first-principles approach. We calculated the
non-linear differential conductance for two structures of a
four-atom wire, and clarified the mode selectivity observed
experimentally as well as the mechanism behind phonon signal
increase with elongation. Further, we considered two extremes of
external mode damping, which lead to the suggestion that local
``heating'' of the wire is significant in the experiment.

We thank the French Embassy in Copenhagen for financial help and
acknowledge stimulating discussions with M.~Paulsson. M.B.~thanks
the CNRS for a ``poste de chercheur associ\'{e},'' and N.L.~is
grateful to the ACI jeunes chercheurs. We also thank the Danish
Center for Scientific Computing (DCSC), the Centre d'Informatique
de l'Enseignement Sup\'erieur (CINES), and the Centre de Calcul de
Midi-Pyr\'enn\'ees (CALMIP) for computational resources.


\bibliographystyle{apsrev}

\end{document}